\documentclass[sigconf,screen, authorversion,nonacm]{acmart}

\usepackage{dcolumn}
\newcolumntype{d}{D{.}{.}{2}}

\usepackage{listings}
\usepackage{tabularx}
\usepackage{todonotes}

\begin{document}

%% The "title" command has an optional parameter,
%% allowing the author to define a "short title" to be used in page headers.
\title{Detecting COPD Through Speech Analysis: A Dataset of Danish Speech and Machine Learning Approach}

%%
%% The "author" command and its associated commands are used to define
%% the authors and their affiliations.
%% Of note is the shared affiliation of the first two authors, and the
%% "authornote" and "authornotemark" commands
%% used to denote shared contribution to the research.
\author{Cuno Sankey-Olsen}
\affiliation{%
  \institution{Aalborg University}
  \department{Department of Computer Science}
  \city{Aalborg}
  \country{Denmark}
}
\email{csanke19@student.aau.dk}

\author{Rasmus Hvass Olesen}
\affiliation{%
  \institution{Aalborg University}
  \department{Department of Computer Science}
  \city{Aalborg}
  \country{Denmark}
}
\email{rolese20@student.aau.dk}

\author{Tobias Oliver Eberhard}
\affiliation{%
  \institution{Aalborg University}
  \department{Department of Computer Science}
  \city{Aalborg}
  \country{Denmark}
}
\email{teberh23@student.aau.dk}

\author{Andreas Triantafyllopoulos}
\affiliation{%
  \institution{TUM University Hospital}
  \department{CHI -- Chair of Health Informatics}
  \city{Munich}
  \country{Germany}
}
\email{andreas.triantafyllopoulos@tum.de}

\author{Björn Schuller}
\affiliation{%
  \institution{TUM University Hospital}
  \department{CHI -- Chair of Health Informatics}
  \city{Munich}
  \country{Germany}
}
\email{schuller@tum.de}

\author{Ilhan Aslan}
\affiliation{%
  \institution{Aalborg University}
  \department{Department of Computer Science}
  \city{Aalborg}
  \country{Denmark}
}
\email{ilas@cs.aau.dk}

%%
%% By default, the full list of authors will be used in the page
%% headers. Often, this list is too long, and will overlap
%% other information printed in the page headers. This command allows
%% the author to define a more concise list
%% of authors' names for this purpose.
\renewcommand{\shortauthors}{}

%%
%% The abstract is a short summary of the work to be presented in the
%% article.
\begin{abstract}

Chronic Obstructive Pulmonary Disease (COPD) is a serious and debilitating disease affecting millions around the world.
Its early detection using non-invasive means could enable preventive interventions that improve quality of life and patient outcomes, with speech recently shown to be a valuable biomarker.
Yet, its validity across different linguistic groups remains to be seen.
To that end, audio data were collected from 96 Danish participants conducting three speech tasks (reading, coughing, sustained vowels).
Half of the participants were diagnosed with different levels of COPD and the other half formed a healthy control group.
Subsequently, we investigated different baseline models using openSMILE features and learnt x-vector embeddings.
We obtained a best accuracy of 67\% using openSMILE features and logistic regression.
Our findings support the potential of speech-based analysis as a non-invasive, remote, and scalable screening tool as part of future COPD healthcare solutions.

\end{abstract}

%%
%% The code below is generated by the tool at http://dl.acm.org/ccs.cfm.
%% Please copy and paste the code instead of the example below.
%%
%
%\begin{CCSXML}
%<ccs2012>
%   <concept>
%       <concept_id>10010147.10010257</concept_id>
%       <concept_desc>Computing methodologies~Machine learning</concept_desc>
%       <concept_significance>500</concept_significance>
%       </concept>
%   <concept>
%       <concept_id>10003120</concept_id>
%    <concept_desc>Human-centered computing</concept_desc>
%       <concept_significance>500</concept_significance>
%       </concept>
% </ccs2012>
%\end{CCSXML}

%\ccsdesc[500]{Computing methodologies~Machine learning}
%\ccsdesc[500]{Human-centered computing}

%%
%% Keywords. The author(s) should pick words that accurately describe
%% the work being presented. Separate the keywords with commas.

%%
%% This command processes the author and affiliation and title
%% information and builds the first part of the formatted document.
\maketitle

\section{Introduction} 
Recent advancements in mobile health and artificial intelligence (AI) have demonstrated the growing potential of machine learning in supporting medical diagnostics and empowering patients beyond traditional clinical environments.
Mobile-first technologies and wearable devices are increasingly being used for private, in-home health assessments and pre-clinical consultations \cite{ExploreSelf, ECG, Li}.
With the combination of sophisticated machine learning techniques and access to large-scale medical datasets, AI-based systems have achieved significant success in fields such as radiology, cardiology, and disease prediction \cite{yuan2025machine, barnes2023machine, adham24}.
Accordingly, speech analysis has been gaining increasing interest in the medical field, as speech and language analysis can assess a wide range of health conditions due to the complexity of speech production in relation to overall health \cite{SpeechForHealth, SpeechAsBlood2022, Triantafyllopoulos23-H4H}. 
This includes both neurological and respiratory health, as even slight changes in these areas can affect a person’s ability to control the vocal apparatus, thereby altering acoustic properties \cite{Hirschberg}.
The unique insights derived from pathological speech changes, combined with the ease of collecting, storing, and programmatically analyzing speech data with AI \cite{Cunningham}, have contributed to the growing role of speech analysis in medicine. 

One area that stands to benefit significantly from these technological advancements is the diagnosis of respiratory diseases and, particularly, Chronic Obstructive Pulmonary Disease (COPD)~\cite{Wolfgang}. Affecting over 390 million people globally, COPD is a progressive and frequently underdiagnosed condition that severely impacts patients’ quality of life.
Research indicates that up to 80\% of COPD cases remain undiagnosed, with many individuals experiencing symptoms for years before receiving a clinical diagnosis \cite{WHO_COPD}.
This trend is also reflected locally in Denmark, where the present study was conducted~\cite{sogaard2016incidence}.
Given the nature of COPD as a respiratory disease and its effect on airflow, voice quality, and phonation~\cite{Triantafyllopoulos22-DBP, Triantafyllopoulos24-SVF, Wolfgang}, the use of speech as a potential biomarker for early screening offers substantial promise.
This approach aligns with broader trends in mobile health and passive sensing technologies, which aim to extend diagnostic capabilities beyond traditional healthcare settings \cite{Bufano23, Theopilus25, Mahmood25} and into everyday environments.

Building on this existing line of work, this study introduces a novel dataset of Danish speech for classifying COPD patients vs healthy controls\footnote{We plan to release anonymized features.}.
We additionally implement a standard pipeline for segmenting the data, extracting features, and training classification models\footnote{Code will be made publicly available.}.
This addresses a notable gap in the present literature, as ensuring the suitability of speech processing for a wide variety of language groups is of paramount importance for scaling early detection tools that rely on speech.

\section{Related work}
This section provides an overview of COPD, including its progression, impact on quality of life, and challenges in diagnosis. Furthermore, it explores the role of machine learning, with a specific focus on its applications in speech analysis for COPD status. 

\subsection{COPD in the Medical Context}
COPD is a progressive respiratory disorder characterized by limited airflow function, primarily caused by chronic bronchitis and damaged alveoli, called emphysema. It affects over 390 million people globally and caused around 3.5 million deaths in 2021, which were roughly 5\% of all deaths globally \cite{WHO_COPD}. 
It is a major public health challenge, with mortality rates expected to rise due to ageing populations and persistent exposure to risk factors such as smoking tobacco, air pollution levels, and occupational hazards \cite{WHO_COPD}. 

COPD is not a static condition but rather a spectrum of disease that is commonly classified into several stages according to the severity of airflow limitation \cite{GOLD2021}. The Global Initiative for Chronic Obstructive Lung Disease (GOLD) guidelines divide COPD into four stages: Mild (GOLD 1), moderate (GOLD 2), severe (GOLD 3), and very severe (GOLD 4). COPD may be almost asymptomatic in the early stage, with patients experiencing only subtle symptoms, such as mild shortness of breath during exertion or the occasional cough \cite{GOLD2021}. As the condition progresses to advanced stages, the limitation of airflow becomes increasingly pronounced. People with moderate to severe disease commonly have a chronic cough, greater volume of sputum expectoration, and shortness of breath that considerably restricts their ability to function normally. This gradual decline illustrates the progressive component of COPD and shows why the initial signs are so frequently ignored or attributed to becoming older or being unfit \cite{GOLD2021}.

\subsubsection{\textbf{Living with COPD and Quality of Life}}
COPD can affect people's lives beyond just a decline in lung function that can be measured. Due to intermittent symptoms in the early stages, many people do not recognize the progression of their condition. Later stages are often associated with substantial physical limitations that negatively affect people's activities of daily living, work, exercise, and socializing \cite{FunctPhysLimitCOPD}. Patients may report chronic breathlessness, fatigue, and frequent respiratory infections, all of which have detrimental effects on physical activity and may lead to psychological comorbidities such as anxiety and depression \cite{GOLD2021}. 
For most people, living with COPD means a life of ongoing adjustment to changing abilities -- modifying daily activities, depending on oxygen therapy, or attending pulmonary rehabilitation programs to preserve the quality of life. The complex relationship between physical symptoms and emotional status underscores the need for comprehensive care that responds to both medical and psychosocial concerns \cite{FunctPhysLimitCOPD}.

\subsubsection{\textbf{The Burden of Undiagnosed COPD}} 
As previously presented in this paper, COPD is a major global health burden, with approximately 60–86\% of cases remaining undiagnosed \cite{GERSHON20181336}. Undiagnosed COPD is linked to poor clinical outcomes, including significantly higher rates of exacerbations, hospitalization, respiratory-related mortality, and impaired quality of life. Delayed diagnosis also accelerates disease progression and the likelihood of increased comorbid conditions \cite{COPDearlyDetection}. 
This source further describes that detection is critical since the treatment of mild-to-moderate COPD, the stage that most individuals have before diagnosis, can slow the progression of the disease. Quitting smoking, medication, and lung rehabilitation have revealed evident benefits in preserving lung function, minimizing exacerbations, and enhancing quality of life. Nevertheless, the paper highlights issues, such as insufficient use of spirometry, overlooking early symptoms, and overemphasis on tobacco hazards, which hinder the timely diagnosis of patients. 
Guidelines now recommend targeted case-finding in high-risk individuals (e.g., adults over the age of 35–40 years with respiratory symptoms, smoking history, or environmental exposure). Technological advances in portable spirometry devices and validated risk assessment questionnaires maximize detection efficiency but are limited by their high cost and low availability. Proactive case-finding strategies, coupled with risk factor evaluation and testing, are required to reduce undiagnosed COPD, improve patient outcomes, and alleviate pressures on healthcare systems \cite{COPDearlyDetection}.

\subsection{COPD detection using speech} 
With the increasing application of AI technology in COPD monitoring and management, modern tools are frequently used for diagnosis, treatment, and post-diagnosis care.
However, speech analysis particularly in the early diagnosis of COPD, remains relatively underutilized \cite{litReviewAICOPD}. 
Our previous work on the classification between exacerbated vs recompensated states in a cohort of COPD patients has aimed to mitigate this gap \cite{Triantafyllopoulos22-DBP, Triantafyllopoulos24-SVF, Wolfgang}.
Specifically, we have investigated the use of both sustained vowels \cite{Triantafyllopoulos24-SVF} and read speech \cite{Triantafyllopoulos22-DBP, Wolfgang} in the classification of COPD patients into \emph{exacerbation} vs \emph{recompensation} -- i.e. classifying between different states of patients.
Other works have focused on distinguishing the speech of COPD patients from that of healthy controls.
These utilise different types of vocalisations, such as breathing~\cite{Ashraf20-VSA, nallanthighal2022detection}, coughing~\citep{Crooks17-CCM}, sustained vowels~\citep{Merker20-DEA}, or read/free speech~\cite{Mohamed14-VCI, Cleres21-LAV, Merker20-DEA, van2021automatic} to distinguish between COPD patients and healthy individuals.
Our own contribution is mostly related to this latter line of work, as we focus on the classification of COPD vs healthy individuals.

\section{Methods}
This section outlines our methodological approach.
It describes our procedure used for participant recruitment, data collection, audio processing, and complementary qualitative interviews.
This process was meant to ensure both technical consistency and ethical compliance while capturing a diverse range of phonatory and respiratory characteristics from individuals with and without COPD. 
he following subsections provide an overview of the dataset, recording protocols, and interview techniques employed during the study.

\subsection{Data Collection} \label{Data_collection}
\begin{figure}[t!]
    \centering
    \includegraphics[width=1\linewidth]{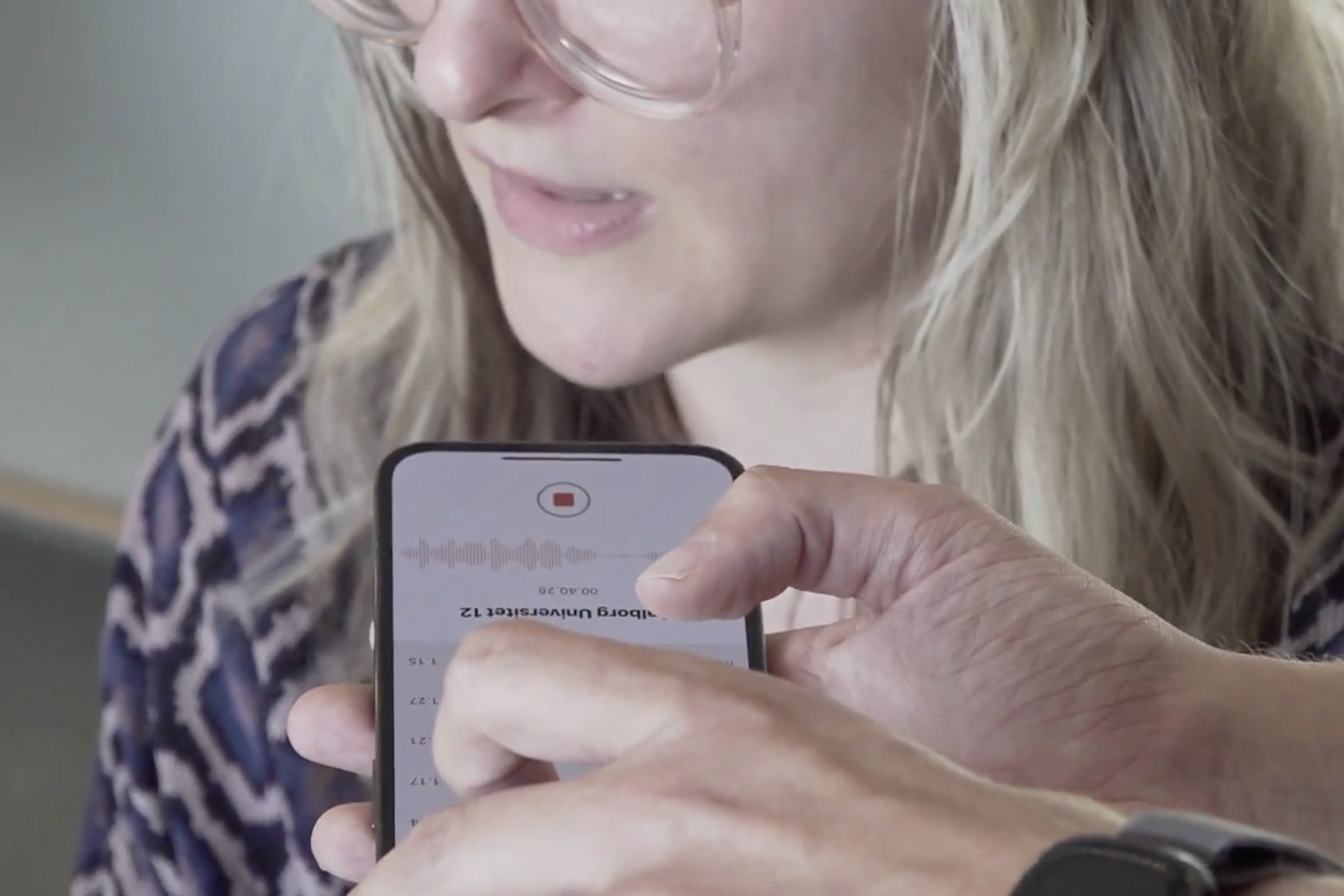}
    \caption{An example of an audio recording}
    \label{fig:audio_rec}
\end{figure}

\begin{figure*}[h]
    \centering
    \includegraphics[width=.9\linewidth]{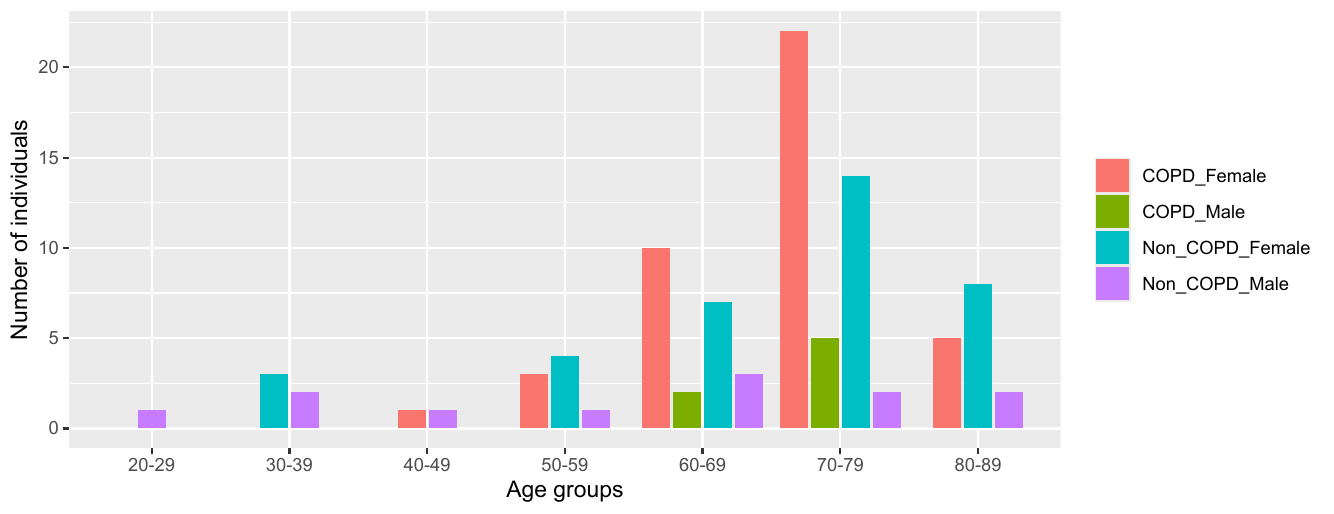}
    \caption{Age Profile of the COPD Participants}
    \label{fig:copd_age}
\end{figure*}

The primary dataset in this study consists of audio recordings from human participants, including both individuals diagnosed with COPD and a control group without any known respiratory conditions. Recordings were collected using microphones on smartphones (iPhone 13 Pro, iPhone 12 Pro, iPhone 10) in quiet, controlled environments; see Figure \ref{fig:audio_rec}. 
Participants were primarily recruited in person through visits to local activity centres, lung cafés, lung choirs, and COPD-specific fitness groups in Aalborg and Aarhus. A smaller number of participants were recruited online through COPD-related Facebook groups, Google Forms, and E-mail. Recordings were conducted either in person or remotely by the participants themselves.
In total, data was collected from 48 participants with varying stages of COPD (GOLD 1 to 4) and 48 participants in the control group, all aged between 26 and 88 years, with each group having a median age of 72. Details are further described in Table \ref{tab:copd_non_copd_stats} and Figure \ref{fig:copd_age}.

\begin{table}[h]
    \centering
    \begin{tabular}{@{}lll@{}}
        \toprule
        \textbf{Statistic}               & \textbf{COPD}         & \textbf{Non-COPD}    \\ \midrule
        \multicolumn{3}{l}{\textbf{COPD GOLD Stages}}              \\ \midrule
        GOLD 1                           & 14 (29.2\%)                    & -                    \\
        GOLD 2                           & 14 (29.2\%)                    & -                    \\
        GOLD 3                           & 14 (29.2\%)                    & -                    \\
        GOLD 4                           & 1 (2.1\%)                     & -                    \\
        GOLD unknown                     & 5 (10.4\%)                     & -                    \\ \midrule
        \multicolumn{3}{l}{\textbf{Gender Distribution}}           \\ \midrule
        Females                          & 41 (age: 49-85)       & 36 (age: 34-88)      \\
        Males                            & 7 (age: 64-74)        & 12 (age: 26-83)      \\ \midrule
        \textbf{Total}                   & \textbf{48}           & \textbf{48}          \\
        \textbf{Median Age}              & \textbf{72}           & \textbf{72}          \\ \bottomrule
    \end{tabular}
    \caption{COPD and Non-COPD Statistics}
    \label{tab:copd_non_copd_stats}
\end{table}

Our data collection protocol was adapted from Mayr et al. \cite{Wolfgang}.
Each participant was asked to perform three specific vocal tasks designed to capture a broad spectrum of respiratory and phonatory characteristics. 

\begin{enumerate}
    \item Sustained vowel pronunciation: Participants continuously pronounced a sequence of Danish vowels – ``A, E, O, Ø, and Å'' -- to assess vocal steadiness and airflow control.
    \item Reading aloud: Participants read the Dutch version of the short ``The North Wind and the Sun'' fable aloud. This test was selected for its standardized structure.
    \item Coughing: Participants were instructed to cough three times into the microphone to capture forced expiratory sounds typically affected by respiratory illness.
\end{enumerate}

\subsection{Dataset Preparation}
The average duration of each participant session was approximately 1–2 minutes. Recordings were manually segmented into individual audio files for each task and stored in WAV format at a sample rate of 44.1 kHz. The only preprocessing method used was silence trimming, using the \textsc{pydub} Python library. In addition, Adobe Audition was used to reduce background noise in a few audio recordings where excessive noise was present. File naming conventions and metadata, such as age, gender, and COPD status, were stored in a structured format to support model training. This structure also included additional, unused metadata, such as the COPD stage and remarks or comments made during the recording sessions. These supplementary data points could be leveraged in future work to train models with alternative focuses or to support more fine-grained analyses. An example can be seen in the snippet in  Listing \ref{metadata} of the JSON metadata file.

\begin{lstlisting}[caption={Metadata for Audio Recordings}, label= {metadata}, captionpos = b]
    {
    "name": "00012",
    "silence_thresh": -40,
    "age": 72,
    "gender": "M",
    "COPD_status": 1,
    "COPD_GOLD": 3,
    "comment": "This person
    could not read the script, 
    so he told a short story instead"
    }
\end{lstlisting}

\subsection{Interview Data}
To complement the quantitative audio data, short, informal, semi-structured interviews were conducted with a subset of participants after they had completed their recordings. 
These followed the qualitative methodology of Brinkmann and Kvale \cite{BrinkmannKvale2018DoingInterviews}, using an interview guide designed to elicit reflections on respiratory health, recent symptoms, and participants’ experiences with the recording process . Given participant demographics, the interviews were conducted in a conversational and approachable manner to promote comfort and openness \cite{Spradley_P_James_1979}. All interviews were audio-recorded and anonymized in accordance with GDPR and institutional ethical guidelines.
Although the interview data was not used to train the machine learning models, it played an important role in validating the interpretation of the audio recordings and enriched the overall dataset. 

The conversations also confirmed several assumptions drawn from existing literature and consultations with medical professionals – particularly the insight that many individuals may unknowingly live with COPD for years before receiving a formal diagnosis. Additionally, the interviews revealed a generally positive attitude among participants, with many expressing appreciation for contributing to research that has the potential to benefit future generations of patients.

\subsection{Experimental setup}
An overview of our model training process is shown in Figure \ref{fig:Model}.
Following the manual segmentation of the data, we subsequently extracted two sets of features.
First, we used openSMILE \cite{openSMILE} to extract the standardized eGeMAPS feature set \cite{eyben2015geneva} that has shown good predictive performance in previous work \cite{Triantafyllopoulos22-DBP, nallanthighal2022detection}.
Additionally, we extracted x-vector embeddings using an \emph{ETDNN} model pre-trained for speaker identification~\cite{Desplanques20-EEC} available in SpeechBrain \cite{speechbrain}.
Finally, we investigated the use of age as one additional extra feature in order to improve performance using demographic information.
Subsequently, we investigated four standard models using nested 5-fold cross-validation, where we ensured that no speakers from the training set were part of the test set in each fold.
We investigated the following models and hyperparameters ($N$ is the number of instances in the training set, $d$ the dimensionality of features, and $\sigma^2$ the variance of the features): Random Forests (number of trees: $\{400, 450, 500\}$; maximum depth: $\{10, 11, 12, 13\}$; minimum samples to split node: $\{12, 13, 14\}$; minimum samples for leaf node: $\{1, 2, .5\cdot N\}$; maximum features to consider for split: $\sqrt{d}$), Support Vector Machines (SVMs; cost factor: $\{.01, .1, .5, 1, 2, 5, 10, 20, 50\}$; kernel: \{linear, polynomial, radial basis function\}; kernel coefficient: $\{.001, .01, .1, 1, 2, 5, \frac{1}{d}, \frac{1}{d \cdot \sigma^2}\}$), Logistic Regression with two solvers, \emph{SAGA}~\cite{defazio2014saga} (penalty: \{\emph{ElasticNet}\}; cost: $\{.01, .03, .04, .05, 1\}$) and \emph{LibLinear}~\cite{fan2008liblinear} (penalty: \{L1-norm, L2-norm\}; cost: $\{.01, .03, .04, .05, 1\}$), and a 4-layered feed-forward neural network (hidden size: $\{256, 128, 64, 32\}$; ReLU, dropout with probability 0.3).
Features are always standardized based on the parameters of the training set in each fold.

\begin{table*}[t!]
\centering
\caption{Performance Metrics for COPD Prediction Models Across Datasets and Feature Sets (0 = non-COPD, 1 = COPD)} \label{MetricsTable}
\footnotesize
\begin{tabularx}{\textwidth}{l l c c c c c c c}
\toprule
\textbf{Model} \hspace{5em} & \textbf{Dataset} & \textbf{Precision (0)} & \textbf{Recall (0)} & \textbf{F1 (0)} & \textbf{Precision (1)} & \textbf{Recall (1)} & \textbf{F1 (1)} & \textbf{Accuracy} \\
\midrule

\multicolumn{9}{l}{\textit{Random Forest}} \\
& x-vectors (w/ age)         & .52 & .57 & .49 & .69 & .66 & .65 & .60 \\
& x-vectors (w/o age)        & .51 & .61 & .50 & .69 & .62 & .62 & .59 \\
& eGeMAPS (w/ age)           & .64 & .59 & .61 & .66 & .70 & .68 & .65 \\
& eGeMAPS (w/o age)          & .64 & .60 & .61 & .66 & .70 & .68 & .65 \\

\addlinespace
\multicolumn{9}{l}{\textit{SVM}} \\
& x-vectors (w/ age)         & .53 & .57 & .53 & .69 & .66 & .66 & .62 \\
& x-vectors (w/o age)        & .53 & .57 & .54 & .69 & .66 & .66 & .62 \\
& eGeMAPS (w/ age)           & .64 & .68 & .65 & .70 & .65 & .67 & .66 \\
& eGeMAPS (w/o age)          & .64 & .66 & .64 & .69 & .67 & .67 & .66 \\

\addlinespace
\multicolumn{9}{l}{\textit{Logistic Regression}} \\
& x-vectors (w/ age)         & .52 & .58 & .53 & .69 & .65 & .66 & .61 \\
& x-vectors (w/o age)        & .52 & .58 & .53 & .69 & .65 & .66 & .61 \\
& eGeMAPS (w/ age)           & .63 & .67 & .65 & .69 & .65 & .67 & .66 \\
& eGeMAPS (w/o age)          & .64 & .66 & .64 & .69 & .66 & .67 & .66 \\

\addlinespace
\multicolumn{9}{l}{\textit{Neural Network}} \\
& x-vectors (w/ age)         & .44 & .54 & .46 & .63 & .54 & .57 & .54 \\
& x-vectors (w/o age)        & .45 & .58 & .49 & .64 & .53 & .56 & .54 \\
& eGeMAPS (w/ age)           & .64 & .64 & .63 & .67 & .65 & .65 & .65 \\
& eGeMAPS (w/o age)          & .64 & .63 & .62 & .68 & .67 &  .66 & .65 \\
\bottomrule
\end{tabularx}
\end{table*}

\section{Results}
This section presents our results.
A total of 16 model results were generated by combining the four machine learning models with the two different feature sets and further combining them with age as an additional feature.
We computed precision, recall, and F1-score for each class separately, as well as their combined accuracy.
The final performance metrics are averaged across all five runs.
A summary of these performance metrics is presented in Table \ref{MetricsTable}\footnote{Class ``0'' corresponds to non-COPD individuals. Class ``1'' refers to those diagnosed with COPD.}.
We present utterance-level performance (i.e., we treat each utterance as one independent trial).
Focusing on accuracy, we note that eGeMAPS provides better results than x-vectors, reaching a top accuracy of $.66$ using Logistic Regression and SVMs.
The inclusion of age as an additional feature does not improve performance further.

\begin{figure}[h]
    \centering
    \includegraphics[width=1\linewidth]{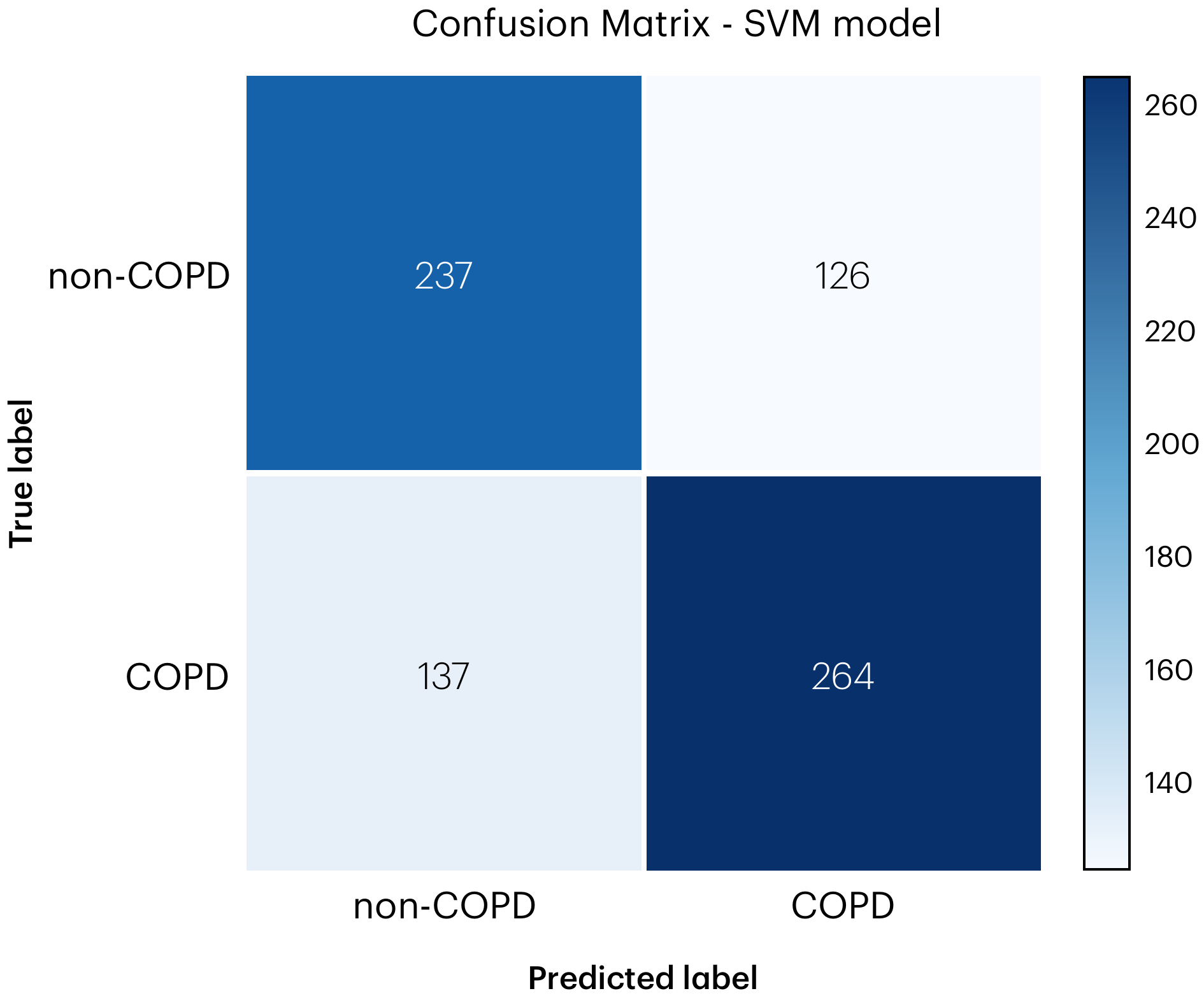}
    \caption{Averaged Confusion Matrix of SVM Model (eGeMAPS).}
    \label{fig:osMatrixSVM}
\end{figure}

Figure \ref{fig:osMatrixSVM} presents the averaged confusion matrix for the SVM model trained with eGeMAPS. Due to its strong performance, this model was selected for detailed illustration through a confusion matrix. The matrix displays the model’s predictions on individual audio chunks across all validation folds. Each chunk was labeled based on the participant's known COPD status (non-COPD and COPD).

\section{Discussion}
This section presents the key findings of the study, examining both the strengths and limitations of the methodological approach and model performance. It aims to contextualize the results presented in the previous section by examining potential sources of bias, data-related challenges, and the implications of demographic variation and model behaviour. Additionally, this section outlines critical factors that may influence the generalizability and robustness of the findings while highlighting avenues for future research and improvement.

\subsection{Limitations}
While the methods employed in this study were carefully designed to ensure consistency and reliability, several limitations emerged during data collection and model development. These limitations may have impacted both the quality of the dataset and the generalizability of the machine learning models. The following subsections outline key challenges and considerations that should be taken into account by future research.

\subsubsection{Collecting Online Recordings}
As mentioned in Section~\ref{Data_collection}, 
recruitment was initiated by publishing a post in several Facebook groups aimed at reaching potential participants. Individuals were given the option to either record themselves or meet in person for assistance with the recording process. The majority chose to record themselves. 
It was initially assumed that a portion of users in these forums might have limited technical proficiency. To address this potential barrier, a short video guide was developed to clearly outline each step required for creating and submitting the recording. However, this did not have any effect on gathering more data from this demographic online.

\subsubsection{Gender Imbalance}
During the collection of the recordings from COPD and non-COPD participants, we attended multiple events where most participants were women, which resulted in an imbalance in our dataset. However, this imbalance in our dataset could also be a result of the prevalence of COPD in females has increased, and the number of females diagnosed with COPD in the United States now outnumbers males \cite{Milne2024Sexdifferences}. Milne at al. \cite{Milne2024Sexdifferences} further argue that a possible reason for this is that females may be more susceptible to the effects of cigarettes compared to males. 
This imbalance limits the generalizability of our models, as the vocal characteristics differ substantially between males and females.
A dataset dominated by females might affect the model's ability to generalize to the broader population, particularly male patients.
Note that for present purposes we consider gender to be equivalent to \emph{sex-assigned-at-birth}, thus ignoring additional confounders due to different social factors and self-perception.

\subsubsection{Unverified Self-Reported Diagnosis}
A significant limitation of this study is the verification of participant COPD status. The classification of individuals into COPD and non-COPD groups was based solely on self-reporting. No medical records, diagnostic test results (such as spirometry), or clinical documentation were obtained to confirm diagnoses. As a result, there is a risk of misclassification, especially among participants in the control group, where undiagnosed COPD may have gone unrecognized. This introduces potential noise into the training data, which could affect model performance.

\subsubsection{Age Limitations}
An age-related limitation was present during the data collection process, as the median age of the participants was 72 years. To ensure the most valid and generalizable results, the control group needed to be of a similar age, aligning with the typical age range at which individuals are commonly diagnosed with COPD \cite{Buchardt}. However, completely excluding younger individuals would reduce the model’s precision if used by a broader demographic.  To address this, a smaller subset of younger participants in their late 20s, 30s, and 40s was also included in the final model.

Another limitation related to age is that COPD symptoms typically begin to manifest in individuals over the age of 40 years \cite{Agarwal}. Additionally, research shows that many people experience symptoms for several years before receiving a formal diagnosis. This presents a potential issue in the dataset -- especially at scale -- as some individuals labeled as `non-COPD' may, in fact, unknowingly exhibit early signs of the disease without yet being diagnosed. As a result, the model may learn from mislabeled data, inadvertently treating early-stage COPD cases as healthy controls. This could reduce the model’s overall precision and hinder its ability to accurately distinguish between healthy and affected individuals, particularly in the early stages of the disease.

\subsubsection{Inclusion of Active Smokers}
This limitation of the study concerns the composition of the control group, which includes individuals who are current or former smokers. While these participants reported no known respiratory conditions, it cannot be ruled out that some might have undiagnosed or early-stage COPD. This introduces a risk of false positives in the classification task, as the model may learn to associate smoking-related vocal characteristics with COPD.
However, excluding smokers entirely from the control group would have introduced another bias, where we would be comparing primarily non-smokers with a COPD group largely composed of participants with a history of smoking. This would risk training the model to distinguish between smoking status rather than the status of their disease. Including smokers in the control group, therefore, reflects a more realistic scenario, where early COPD is often undiagnosed, but also underlines the need for awareness in interpreting borderline cases.

\subsubsection{Sustained Vowels}
Another limitation relates to the variability in how participants performed the sustained vowel pronunciation task. The purpose of this task was to capture long, steady vowel sounds that could help highlight vocal or respiratory differences between individuals with and without COPD. However, several participants did not follow the instructions as intended. In some cases, vowels were pronounced quickly or with little effort to sustain the sound, while others varied in how long or clearly they vocalized the vowels. Furthermore, some participants were either uncomfortable with or unable to read aloud. In such cases, they instead shared a short anecdote or a personal story. 
This inconsistency in task execution may have introduced differences in the recordings that are unrelated to COPD, which can affect how the machine learning models interpret and learn from the data. When the input quality varies from one participant to another, it becomes more difficult for the model to focus on the disease-related patterns. Although this kind of variation reflects real-world user behavior, it also highlights the importance of providing clearer instructions and possibly excluding recordings that do not meet the expected task format in future studies.

\subsubsection{Post-Processing of Audio}
During the post-processing of the collected audio recordings, several samples contained background noise or music that needed to be removed 
to prevent confusion for the models. However, even when applied carefully, noise reduction tools can unintentionally eliminate important speech features. This may result in the loss of relevant information critical for accurately detecting patterns associated with COPD, potentially degrading the model's performance.

\subsection{Model performance}
Having discussed the limitations our methodology, we now turn to the performance and interpretation of our model.
This section evaluates and compares the performance of the various machine learning models trained to classify COPD from voice recordings. The discussion is structured around model types, feature sets, and the impact of demographic information.\\
Reviewing the results of the trained models in Table~\ref{MetricsTable}, one of the first notable observations is the overall consistency in performance across most models. Interestingly, the feed-forward neural network exhibits the lowest accuracy among all evaluated models. Several factors likely contribute to this outcome -- particularly when using x-vectors.
Neural networks generally require substantially larger datasets to generalize effectively and avoid overfitting, which we assume is the main reason for its lower performance here.

In addition to data limitations, the architectural design of neural networks may also influence performance. Unlike traditional machine learning models, neural networks are often considered `black boxes,' as the internal processes by which inputs are transformed into outputs are not easily interpretable. This lack of transparency complicates efforts to diagnose performance issues, which may stem from data scarcity, overfitting, or suboptimal model design—such as the number and type of layers, the number of nodes per layer, or the choice of activation functions. It is, therefore, plausible that the neural network implemented in this study was not optimally configured for the relatively small dataset and high-dimensional feature representation, further contributing to its lower accuracy.

When evaluating the top-performing model triads – Random Forest, SVM, and Logistic Regression trained with the eGeMAPS feature set – all demonstrate accuracy within the range of 64–66\%. Notably, the SVM model trained without incorporating age metadata, achieves a well-balanced recall: 67\% for COPD cases and 66\% for non-COPD cases. Logistic Regression shows a similar trend, while Random Forest tends to prioritize recall for COPD cases (72\%) at the cost of reduced performance for non-COPD cases. These differences can be attributed to the fundamental mechanics of each algorithm. Random Forest, an ensemble method based on decision trees, is generally robust against overfitting and does not require feature scaling. However, this robustness may come at the expense of nuanced predictions in smaller datasets, especially for underrepresented classes. Like Neural Networks, Random Forest models typically perform better with larger datasets, where their ensemble nature can leverage greater variance. Conversely, SVM and Logistic Regression may perform adequately on smaller datasets but are more sensitive to issues such as class imbalance and the absence of detailed feature scaling. While these models are also more prone to overfitting under these conditions, the use of validation techniques such as stratified cross-validation helps mitigate these risks by preserving class distributions across training and test folds.

\subsubsection{\textbf{Proof of Concept}}
While the current findings highlight promising avenues for classifying COPD from voice recordings, the modest accuracy and class imbalance underscore limitations in the existing dataset. A pronounced gender skew and a limited overall sample size constrain both the generalizability and robustness of the results. Future research should prioritize expanding the dataset -- particularly by including more male participants and increasing the total number of samples. 
Furthermore, incorporating domain knowledge for feature selection or exploring multimodal data fusion (e.g., combining audio with sensor or questionnaire data) could further enhance model performance.
An additional opportunity lies in utilizing the collected COPD-GOLD status data (see listing \ref{metadata}). Training models specifically on early-stage cases (GOLD 1–2), combined with a larger and more diverse dataset, could potentially enhance the ability to detect early signs of COPD more effectively. Another promising direction would be to train models capable of distinguishing between GOLD stages. Such a system could be valuable for monitoring disease progression in diagnosed patients, assessing whether their condition is improving or worsening – potentially estimating lung function remotely without requiring a hospital visit.

These findings demonstrate the feasibility of using voice features to classify COPD, particularly when combined with metadata such as age. While the models are not yet robust enough for clinical deployment, the observed performance, especially in female participants, provides compelling proof of concept. Future work with larger, more balanced datasets and additional feature engineering could further improve predictive accuracy and generalizability.                                                                      

\subsubsection{Future Outlook}
While the current findings highlight promising avenues for classifying COPD from voice recordings, the modest accuracy and class imbalance underscore limitations in the existing dataset. A pronounced gender skew and a limited overall sample size constrain both the generalizability and robustness of the results. Future research should prioritize expanding the dataset -- particularly by including more male participants and increasing the total number of samples. 
Furthermore, incorporating domain knowledge for feature selection or exploring multimodal data fusion (e.g., combining audio with sensor or questionnaire data) could further enhance model performance.
An additional opportunity lies in utilizing the collected COPD-GOLD status data (see listing \ref{metadata}). Training models specifically on early-stage cases (GOLD 1–2), combined with a larger and more diverse dataset, could potentially enhance the ability to detect early signs of COPD more effectively. Another promising direction would be to train models capable of distinguishing between GOLD stages. Such a system could be valuable for monitoring disease progression in diagnosed patients, assessing whether their condition is improving or worsening – potentially estimating lung function remotely without requiring a hospital visit.

The prospect of using advanced speech analysis techniques on mobile devices promises novel and context-aware remote healthcare solutions. Context-awareness on mobile devices is an established interaction paradigm \cite{gellersen2002multi}, which has found its application early in conversational, multimodal, and multilingual interfaces on mobile devices \cite{compass2008}. Clearly, reducing the increasing burden on the healthcare sector with health-aware technologies is an important goal. However, it is also important to acknowledge the increase in dependability on such devices and infrastructures for delivering basic health services. Furthermore, while functions to perform health analysis from speech could be embedded easily in  consumer devices and mobile applications there are important ethical issues to consider. For example, who should be allowed to perform such an health analysis and who  should have access to the results? Overall, it is promising that challenges and opportunities for research at the intersection of HCI and health is being identified and addressed \cite{blandford2020opportunities}. Furthermore, IxD and HCI researchers are following the call to move towards positive computing \cite{CalvoPeters2012} and well-being promoting designs (e.g., \cite{Bittner2019, compliment2023}). While the future is not defined yet, health solutions and how we deliver healthcare is undergoing a transformation. In future research, we expect to see an increasing integration of health analysis technologies, such as speech analysis with emerging interaction paradigms, especially applications of large language models and extended reality.  

\section{Conclusion}

This study explored the feasibility of utilizing voice recordings and machine learning techniques to detect COPD. Audio data were collected from individuals diagnosed with COPD as well as from a control group without known respiratory conditions. Four classification models — Logistic Regression, Support Vector Machine (SVM), Random Forest, and a Neural Network were trained on acoustic features extracted using both a classic, expert feature set (eGeMAPS) and learnt x-vectors from a pretrained model.
The findings demonstrate that several models, particularly those utilizing eGeMAPS, are capable of distinguishing between COPD and non-COPD cases with moderate accuracy, recall, and F1-scores. While performance varied, the results show the promise of leveraging voice analysis in COPD screening. In particular, the SVM and Random Forest models showed consistently balanced classification outcomes across multiple experimental setups.
Despite discussed limitation, our work makes a meaningful contribution to the emerging field of speech analysis by demonstrating the potential of low-cost, non-invasive diagnostic tools based on voice. This approach holds promise for scalable, at-home screening of COPD. Future work should aim to significantly expand the dataset, especially by including more male participants and clinically validated cases, and explore whether machine learning models can not only detect COPD but also assess its severity or progression.

%\section{Appendices}
%%
%% The acknowledgments section is defined using the "acks" environment
%% (and NOT an unnumbered section). This ensures the proper
%% identification of the section in the article metadata, and the
%% consistent spelling of the heading.

\begin{acks}
We would like to express our sincere gratitude to all those who contributed to the success of this project.
Special thanks go to Lungeforeningen and Nordkraft Sundhedscenter Aalborg for their support in helping us recruit participants and collect valuable recordings from individuals living with COPD. We also appreciate the cooperation of Aarhus Lung Choir and Aarhus Music School during our data collection process. We are particularly thankful to Ulla Møller Weinreich from Lungemedicinsk Afdeling for her generous help in facilitating access to patients and connecting us with relevant contacts.
\end{acks}

\balance

%%
%% The next two lines define the bibliography style to be used, and
%% the bibliography file.
\bibliographystyle{ACM-Reference-Format}
\bibliography{References}

%%% -*-BibTeX-*-
%%% Do NOT edit. File created by BibTeX with style
%%% ACM-Reference-Format-Journals [18-Jan-2012].

\begin{thebibliography}{47}

%%% ====================================================================
%%% NOTE TO THE USER: you can override these defaults by providing
%%% customized versions of any of these macros before the \bibliography
%%% command.  Each of them MUST provide its own final punctuation,
%%% except for \shownote{}, \showDOI{}, and \showURL{}.  The latter two
%%% do not use final punctuation, in order to avoid confusing it with
%%% the Web address.
%%%
%%% To suppress output of a particular field, define its macro to expand
%%% to an empty string, or better, \unskip, like this:
%%%
%%% \newcommand{\showDOI}[1]{\unskip}   % LaTeX syntax
%%%
%%% \def \showDOI #1{\unskip}           % plain TeX syntax
%%%
%%% ====================================================================

\ifx \showCODEN    \undefined \def \showCODEN     #1{\unskip}     \fi
\ifx \showDOI      \undefined \def \showDOI       #1{#1}\fi
\ifx \showISBNx    \undefined \def \showISBNx     #1{\unskip}     \fi
\ifx \showISBNxiii \undefined \def \showISBNxiii  #1{\unskip}     \fi
\ifx \showISSN     \undefined \def \showISSN      #1{\unskip}     \fi
\ifx \showLCCN     \undefined \def \showLCCN      #1{\unskip}     \fi
\ifx \shownote     \undefined \def \shownote      #1{#1}          \fi
\ifx \showarticletitle \undefined \def \showarticletitle #1{#1}   \fi
\ifx \showURL      \undefined \def \showURL       {\relax}        \fi
% The following commands are used for tagged output and should be
% invisible to TeX
\providecommand\bibfield[2]{#2}
\providecommand\bibinfo[2]{#2}
\providecommand\natexlab[1]{#1}
\providecommand\showeprint[2][]{arXiv:#2}

\bibitem[Ashraf et~al\mbox{.}(2020)]%
        {Ashraf20-VSA}
\bibfield{author}{\bibinfo{person}{Obaid Ashraf}, \bibinfo{person}{Erica Rabold}, \bibinfo{person}{Katrina Schlichtkrull}, \bibinfo{person}{Anil Singh}, \bibinfo{person}{Satya Venneti}, \bibinfo{person}{Mir Mohammed Daanish~Ali Khan}, \bibinfo{person}{Rajat Kulshreshtha}, {and} \bibinfo{person}{Prakhar~Pradeep Naval}.} \bibinfo{year}{2020}\natexlab{}.
\newblock \showarticletitle{Voice-based screening and monitoring of chronic respiratory conditions}.
\newblock \bibinfo{journal}{\emph{Chest}} \bibinfo{volume}{158}, \bibinfo{number}{4} (\bibinfo{year}{2020}), \bibinfo{pages}{A1687}.
\newblock


\bibitem[Aslan et~al\mbox{.}(2023)]%
        {compliment2023}
\bibfield{author}{\bibinfo{person}{Ilhan Aslan}, \bibinfo{person}{Dominik Neu}, \bibinfo{person}{Daniela Neupert}, \bibinfo{person}{Stefan Grafberger}, \bibinfo{person}{Nico Weise}, \bibinfo{person}{Pascal Pfeil}, {and} \bibinfo{person}{Maximilian Kuschewski}.} \bibinfo{year}{2023}\natexlab{}.
\newblock \showarticletitle{How to Compliment a Human-Designing Affective and Well-being Promoting Conversational Things}.
\newblock \bibinfo{journal}{\emph{Interaction Design and Architecture (s) Journal}}  \bibinfo{volume}{58} (\bibinfo{year}{2023}), \bibinfo{pages}{157--184}.
\newblock
\urldef\tempurl%
\url{https://doi.org/10.55612/s-5002-058-007}
\showDOI{\tempurl}


\bibitem[Aslan et~al\mbox{.}(2005)]%
        {compass2008}
\bibfield{author}{\bibinfo{person}{Ilhan Aslan}, \bibinfo{person}{Feiyu Xu}, \bibinfo{person}{Hans Uszkoreit}, \bibinfo{person}{Antonio Kr{\"u}ger}, {and} \bibinfo{person}{J{\"o}rg Steffen}.} \bibinfo{year}{2005}\natexlab{}.
\newblock \showarticletitle{COMPASS2008: Multimodal, multilingual and crosslingual interaction for mobile tourist guide applications}. In \bibinfo{booktitle}{\emph{International Conference on Intelligent Technologies for Interactive Entertainment}}. Springer, \bibinfo{pages}{3--12}.
\newblock
\urldef\tempurl%
\url{https://doi.org/10.1007/11590323_1}
\showDOI{\tempurl}


\bibitem[Barnes et~al\mbox{.}(2023)]%
        {barnes2023machine}
\bibfield{author}{\bibinfo{person}{Hayley Barnes}, \bibinfo{person}{Stephen~M Humphries}, \bibinfo{person}{Peter~M George}, \bibinfo{person}{Deborah Assayag}, \bibinfo{person}{Ian Glaspole}, \bibinfo{person}{John~A Mackintosh}, \bibinfo{person}{Tamera~J Corte}, \bibinfo{person}{Marilyn Glassberg}, \bibinfo{person}{Kerri~A Johannson}, \bibinfo{person}{Lucio Calandriello}, {et~al\mbox{.}}} \bibinfo{year}{2023}\natexlab{}.
\newblock \showarticletitle{Machine learning in radiology: the new frontier in interstitial lung diseases}.
\newblock \bibinfo{journal}{\emph{The Lancet Digital Health}} \bibinfo{volume}{5}, \bibinfo{number}{1} (\bibinfo{year}{2023}), \bibinfo{pages}{e41--e50}.
\newblock


\bibitem[Bian et~al\mbox{.}(2024)]%
        {litReviewAICOPD}
\bibfield{author}{\bibinfo{person}{Hupo Bian}, \bibinfo{person}{Shaoqi Zhu}, \bibinfo{person}{Yonghua Zhang}, \bibinfo{person}{Qiang Fei}, \bibinfo{person}{Xiuhua Peng}, \bibinfo{person}{Zanhui Jin}, \bibinfo{person}{Tianxiang Zhou}, {and} \bibinfo{person}{Hongxing Zhao}.} \bibinfo{year}{2024}\natexlab{}.
\newblock \showarticletitle{Artificial Intelligence in Chronic Obstructive Pulmonary Disease: Research Status, Trends, and Future Directions -- A Bibliometric Analysis from 2009 to 2023}.
\newblock \bibinfo{journal}{\emph{International Journal of Chronic Obstructive Pulmonary Disease}}  \bibinfo{volume}{19} (\bibinfo{year}{2024}), \bibinfo{pages}{1849--1864}.
\newblock
\urldef\tempurl%
\url{https://doi.org/10.2147/COPD.S474402}
\showDOI{\tempurl}


\bibitem[Bittner et~al\mbox{.}(2019)]%
        {Bittner2019}
\bibfield{author}{\bibinfo{person}{Bj\"{o}rn Bittner}, \bibinfo{person}{Ilhan Aslan}, \bibinfo{person}{Chi~Tai Dang}, {and} \bibinfo{person}{Elisabeth Andr\'{e}}.} \bibinfo{year}{2019}\natexlab{}.
\newblock \showarticletitle{Of Smarthomes, IoT Plants, and Implicit Interaction Design}. In \bibinfo{booktitle}{\emph{Proceedings of the Thirteenth International Conference on Tangible, Embedded, and Embodied Interaction}} (Tempe, Arizona, USA) \emph{(\bibinfo{series}{TEI '19})}. \bibinfo{publisher}{Association for Computing Machinery}, \bibinfo{address}{New York, NY, USA}, \bibinfo{pages}{145–154}.
\newblock
\showISBNx{9781450361965}
\urldef\tempurl%
\url{https://doi.org/10.1145/3294109.3295618}
\showDOI{\tempurl}


\bibitem[Blandford et~al\mbox{.}(2020)]%
        {blandford2020opportunities}
\bibfield{author}{\bibinfo{person}{Ann Blandford}, \bibinfo{person}{Janet Wesson}, \bibinfo{person}{René Amalberti}, \bibinfo{person}{Raed AlHazme}, {and} \bibinfo{person}{Ragad Allwihan}.} \bibinfo{year}{2020}\natexlab{}.
\newblock \showarticletitle{Opportunities and challenges for telehealth within, and beyond, a pandemic}.
\newblock \bibinfo{journal}{\emph{The Lancet Global Health}} \bibinfo{volume}{8}, \bibinfo{number}{11} (\bibinfo{year}{2020}), \bibinfo{pages}{e1364--e1365}.
\newblock
\showISSN{2214-109X}
\urldef\tempurl%
\url{https://doi.org/10.1016/S2214-109X(20)30362-4}
\showDOI{\tempurl}


\bibitem[Brinkmann and Kvale(2018)]%
        {BrinkmannKvale2018DoingInterviews}
\bibfield{author}{\bibinfo{person}{Svend Brinkmann} {and} \bibinfo{person}{Steinar Kvale}.} \bibinfo{year}{2018}\natexlab{}.
\newblock \bibinfo{booktitle}{\emph{Doing Interviews}}.
\newblock \bibinfo{publisher}{SAGE Publications Ltd}, \bibinfo{address}{London :}.
\newblock
\showISBNx{9781526426093}
\urldef\tempurl%
\url{http://digital.casalini.it/9781526426093}
\showURL{%
\tempurl}


\bibitem[Buchardt et~al\mbox{.}(2024)]%
        {Buchardt}
\bibfield{author}{\bibinfo{person}{Sebastian Tristan~Ehlert Buchardt}, \bibinfo{person}{Ulla~Møller Weinreich}, \bibinfo{person}{Filip~Lyng Lindgren}, \bibinfo{person}{Marie~Dam Lauridsen}, \bibinfo{person}{Johanne~Hermann Karlsen}, \bibinfo{person}{Kristian Kragholm}, \bibinfo{person}{Christian Torp-Pedersen}, {and} \bibinfo{person}{Peter~Ascanius Jacobsen}.} \bibinfo{year}{2024}\natexlab{}.
\newblock \showarticletitle{Patient characteristics and mortality across diagnostic settings in COPD}.
\newblock \bibinfo{journal}{\emph{Respiratory Medicine}}  \bibinfo{volume}{234} (\bibinfo{year}{2024}), \bibinfo{pages}{107843}.
\newblock
\showISSN{0954-6111}
\urldef\tempurl%
\url{https://doi.org/10.1016/j.rmed.2024.107843}
\showDOI{\tempurl}


\bibitem[Bufano et~al\mbox{.}(2023)]%
        {Bufano23}
\bibfield{author}{\bibinfo{person}{Pasquale Bufano}, \bibinfo{person}{Marco Laurino}, \bibinfo{person}{Sara Said}, \bibinfo{person}{Alessandro Tognetti}, {and} \bibinfo{person}{Danilo Menicucci}.} \bibinfo{year}{2023}\natexlab{}.
\newblock \showarticletitle{Digital Phenotyping for Monitoring Mental Disorders: Systematic Review}.
\newblock \bibinfo{journal}{\emph{J Med Internet Res}}  \bibinfo{volume}{25} (\bibinfo{date}{13 Dec} \bibinfo{year}{2023}), \bibinfo{pages}{e46778}.
\newblock
\showISSN{1438-8871}
\urldef\tempurl%
\url{https://doi.org/10.2196/46778}
\showDOI{\tempurl}


\bibitem[Calvo and Peters(2012)]%
        {CalvoPeters2012}
\bibfield{author}{\bibinfo{person}{Rafael~A. Calvo} {and} \bibinfo{person}{Dorian Peters}.} \bibinfo{year}{2012}\natexlab{}.
\newblock \showarticletitle{Positive computing: technology for a wiser world}.
\newblock \bibinfo{journal}{\emph{Interactions}} \bibinfo{volume}{19}, \bibinfo{number}{4} (\bibinfo{date}{July} \bibinfo{year}{2012}), \bibinfo{pages}{28–31}.
\newblock
\showISSN{1072-5520}
\urldef\tempurl%
\url{https://doi.org/10.1145/2212877.2212886}
\showDOI{\tempurl}


\bibitem[Cleres et~al\mbox{.}(2021)]%
        {Cleres21-LAV}
\bibfield{author}{\bibinfo{person}{David Cleres}, \bibinfo{person}{Frank Rassouli}, \bibinfo{person}{Martin Brutsche}, \bibinfo{person}{Tobias Kowatsch}, {and} \bibinfo{person}{Filipe Barata}.} \bibinfo{year}{2021}\natexlab{}.
\newblock \showarticletitle{Lena: a Voice-Based Conversational Agent for Remote Patient Monitoring in Chronic Obstructive Pulmonary Disease}. In \bibinfo{booktitle}{\emph{Proc.\ ACM Conference on Intelligent User Interfaces}}.
\newblock


\bibitem[Crooks et~al\mbox{.}(2017)]%
        {Crooks17-CCM}
\bibfield{author}{\bibinfo{person}{Michael~G Crooks}, \bibinfo{person}{Albertus Den~Brinker}, \bibinfo{person}{Yvette Hayman}, \bibinfo{person}{James~D Williamson}, \bibinfo{person}{Andrew Innes}, \bibinfo{person}{Caroline~E Wright}, \bibinfo{person}{Peter Hill}, {and} \bibinfo{person}{Alyn~H Morice}.} \bibinfo{year}{2017}\natexlab{}.
\newblock \showarticletitle{Continuous cough monitoring using ambient sound recording during convalescence from a COPD exacerbation}.
\newblock \bibinfo{journal}{\emph{Lung}} \bibinfo{volume}{195}, \bibinfo{number}{3} (\bibinfo{year}{2017}), \bibinfo{pages}{289--294}.
\newblock


\bibitem[Cummins et~al\mbox{.}(2018)]%
        {SpeechForHealth}
\bibfield{author}{\bibinfo{person}{Nicholas Cummins}, \bibinfo{person}{Alice Baird}, {and} \bibinfo{person}{Bjoern~W Schuller}.} \bibinfo{year}{2018}\natexlab{}.
\newblock \showarticletitle{Speech analysis for health: Current state-of-the-art and the increasing impact of deep learning}.
\newblock \bibinfo{journal}{\emph{Methods}}  \bibinfo{volume}{151} (\bibinfo{year}{2018}), \bibinfo{pages}{41--54}.
\newblock


\bibitem[Cunningham et~al\mbox{.}(2017)]%
        {Cunningham}
\bibfield{author}{\bibinfo{person}{Stuart Cunningham}, \bibinfo{person}{Phil Green}, \bibinfo{person}{Heidi Christensen}, \bibinfo{person}{José~Joaquín Atria}, \bibinfo{person}{André Coy}, \bibinfo{person}{Massimiliano Malavasi}, \bibinfo{person}{Lorenzo Desideri}, {and} \bibinfo{person}{Frank Rudzicz}.} \bibinfo{year}{2017}\natexlab{}.
\newblock \showarticletitle{Cloud-Based Speech Technology for Assistive Technology Applications (CloudCAST)}.
\newblock \bibinfo{journal}{\emph{Studies in Health Technology and Informatics}}  \bibinfo{volume}{242} (\bibinfo{year}{2017}), \bibinfo{pages}{322 – 329}.
\newblock
\urldef\tempurl%
\url{https://doi.org/10.3233/978-1-61499-798-6-322}
\showDOI{\tempurl}
\newblock
\shownote{Cited by: 6; All Open Access, Green Open Access}.


\bibitem[Defazio et~al\mbox{.}(2014)]%
        {defazio2014saga}
\bibfield{author}{\bibinfo{person}{Aaron Defazio}, \bibinfo{person}{Francis Bach}, {and} \bibinfo{person}{Simon Lacoste-Julien}.} \bibinfo{year}{2014}\natexlab{}.
\newblock \showarticletitle{SAGA: A fast incremental gradient method with support for non-strongly convex composite objectives}.
\newblock \bibinfo{journal}{\emph{Advances in neural information processing systems}}  \bibinfo{volume}{27} (\bibinfo{year}{2014}).
\newblock


\bibitem[Desplanques et~al\mbox{.}(2020)]%
        {Desplanques20-EEC}
\bibfield{author}{\bibinfo{person}{Brecht Desplanques}, \bibinfo{person}{Jenthe Thienpondt}, {and} \bibinfo{person}{Kris Demuynck}.} \bibinfo{year}{2020}\natexlab{}.
\newblock \showarticletitle{{ECAPA-TDNN: Emphasized channel attention, propagation and aggregation in tdnn based speaker verification}}. In \bibinfo{booktitle}{\emph{Proceedings of INTERSPEECH}}. \bibinfo{address}{Shanghai, China}, \bibinfo{pages}{3830--3834}.
\newblock


\bibitem[{El Sherbini} et~al\mbox{.}(2024)]%
        {adham24}
\bibfield{author}{\bibinfo{person}{Adham {El Sherbini}}, \bibinfo{person}{Robert~S. Rosenson}, \bibinfo{person}{Mahmoud {Al Rifai}}, \bibinfo{person}{Hafeez Ul~Hassan Virk}, \bibinfo{person}{Zhen Wang}, \bibinfo{person}{Salim Virani}, \bibinfo{person}{Benjamin~S. Glicksberg}, \bibinfo{person}{Carl~J. Lavie}, {and} \bibinfo{person}{Chayakrit Krittanawong}.} \bibinfo{year}{2024}\natexlab{}.
\newblock \showarticletitle{Artificial intelligence in preventive cardiology}.
\newblock \bibinfo{journal}{\emph{Progress in Cardiovascular Diseases}}  \bibinfo{volume}{84} (\bibinfo{year}{2024}), \bibinfo{pages}{76--89}.
\newblock
\showISSN{0033-0620}
\urldef\tempurl%
\url{https://doi.org/10.1016/j.pcad.2024.03.002}
\showDOI{\tempurl}
\newblock
\shownote{Preventive Medicine and Cardiac CT}.


\bibitem[Eyben et~al\mbox{.}(2015)]%
        {eyben2015geneva}
\bibfield{author}{\bibinfo{person}{Florian Eyben}, \bibinfo{person}{Klaus~R Scherer}, \bibinfo{person}{Bj{\"o}rn~W Schuller}, \bibinfo{person}{Johan Sundberg}, \bibinfo{person}{Elisabeth Andr{\'e}}, \bibinfo{person}{Carlos Busso}, \bibinfo{person}{Laurence~Y Devillers}, \bibinfo{person}{Julien Epps}, \bibinfo{person}{Petri Laukka}, \bibinfo{person}{Shrikanth~S Narayanan}, {et~al\mbox{.}}} \bibinfo{year}{2015}\natexlab{}.
\newblock \showarticletitle{The Geneva minimalistic acoustic parameter set (GeMAPS) for voice research and affective computing}.
\newblock \bibinfo{journal}{\emph{IEEE transactions on affective computing}} \bibinfo{volume}{7}, \bibinfo{number}{2} (\bibinfo{year}{2015}), \bibinfo{pages}{190--202}.
\newblock


\bibitem[Eyben et~al\mbox{.}(2010)]%
        {openSMILE}
\bibfield{author}{\bibinfo{person}{Florian Eyben}, \bibinfo{person}{Martin W{\"o}llmer}, {and} \bibinfo{person}{Bj{\"o}rn Schuller}.} \bibinfo{year}{2010}\natexlab{}.
\newblock \showarticletitle{Opensmile: the munich versatile and fast open-source audio feature extractor}. In \bibinfo{booktitle}{\emph{Proceedings of the 18th ACM international conference on Multimedia}}. \bibinfo{pages}{1459--1462}.
\newblock


\bibitem[Fan et~al\mbox{.}(2008)]%
        {fan2008liblinear}
\bibfield{author}{\bibinfo{person}{Rong-En Fan}, \bibinfo{person}{Kai-Wei Chang}, \bibinfo{person}{Cho-Jui Hsieh}, \bibinfo{person}{Xiang-Rui Wang}, {and} \bibinfo{person}{Chih-Jen Lin}.} \bibinfo{year}{2008}\natexlab{}.
\newblock \showarticletitle{LIBLINEAR: A library for large linear classification}.
\newblock \bibinfo{journal}{\emph{the Journal of machine Learning research}}  \bibinfo{volume}{9} (\bibinfo{year}{2008}), \bibinfo{pages}{1871--1874}.
\newblock


\bibitem[Gellersen et~al\mbox{.}(2002)]%
        {gellersen2002multi}
\bibfield{author}{\bibinfo{person}{Hans~W Gellersen}, \bibinfo{person}{Albrecht Schmidt}, {and} \bibinfo{person}{Michael Beigl}.} \bibinfo{year}{2002}\natexlab{}.
\newblock \showarticletitle{Multi-sensor context-awareness in mobile devices and smart artifacts}.
\newblock \bibinfo{journal}{\emph{Mobile Networks and Applications}} \bibinfo{volume}{7}, \bibinfo{number}{5} (\bibinfo{year}{2002}), \bibinfo{pages}{341--351}.
\newblock
\urldef\tempurl%
\url{https://doi.org/10.1023/A:1016587515822}
\showDOI{\tempurl}


\bibitem[Gershon et~al\mbox{.}(2018)]%
        {GERSHON20181336}
\bibfield{author}{\bibinfo{person}{Andrea~S. Gershon}, \bibinfo{person}{Deva Thiruchelvam}, \bibinfo{person}{Kenneth~R. Chapman}, \bibinfo{person}{Shawn~D. Aaron}, \bibinfo{person}{Matthew~B. Stanbrook}, \bibinfo{person}{Jean Bourbeau}, \bibinfo{person}{Wan Tan}, {and} \bibinfo{person}{Teresa To}.} \bibinfo{year}{2018}\natexlab{}.
\newblock \showarticletitle{Health Services Burden of Undiagnosed and Overdiagnosed COPD}.
\newblock \bibinfo{journal}{\emph{Chest}} \bibinfo{volume}{153}, \bibinfo{number}{6} (\bibinfo{year}{2018}), \bibinfo{pages}{1336--1346}.
\newblock
\showISSN{0012-3692}
\urldef\tempurl%
\url{https://doi.org/10.1016/j.chest.2018.01.038}
\showDOI{\tempurl}


\bibitem[Halpin et~al\mbox{.}(2021)]%
        {GOLD2021}
\bibfield{author}{\bibinfo{person}{David~MG Halpin}, \bibinfo{person}{Gerard~J Criner}, \bibinfo{person}{Alberto Papi}, \bibinfo{person}{Dave Singh}, \bibinfo{person}{Antonio Anzueto}, \bibinfo{person}{Fernando~J Martinez}, \bibinfo{person}{Alvar~A Agusti}, {and} \bibinfo{person}{Claus~F Vogelmeier}.} \bibinfo{year}{2021}\natexlab{}.
\newblock \showarticletitle{Global initiative for the diagnosis, management, and prevention of chronic obstructive lung disease. The 2020 GOLD science committee report on COVID-19 and chronic obstructive pulmonary disease}.
\newblock \bibinfo{journal}{\emph{American journal of respiratory and critical care medicine}} \bibinfo{volume}{203}, \bibinfo{number}{1} (\bibinfo{year}{2021}), \bibinfo{pages}{24--36}.
\newblock


\bibitem[Julia~Hirschberg(2010)]%
        {Hirschberg}
\bibfield{author}{\bibinfo{person}{Noémie~Elhadad Julia~Hirschberg, Anna~Hjalmarsson}.} \bibinfo{year}{2010}\natexlab{}.
\newblock \showarticletitle{``You’re as Sick as You Sound'': Using Computational Approaches for Modeling Speaker State to Gauge Illness and Recovery}.
\newblock \bibinfo{journal}{\emph{Advances in Speech Recognition: Mobile Environments, Call Centers and Clinics}} (\bibinfo{year}{2010}), \bibinfo{pages}{305--322}.
\newblock


\bibitem[Li et~al\mbox{.}(2025)]%
        {Li}
\bibfield{author}{\bibinfo{person}{Brenna Li}, \bibinfo{person}{Saba Tauseef}, \bibinfo{person}{Khai~N. Truong}, {and} \bibinfo{person}{Alex Mariakakis}.} \bibinfo{year}{2025}\natexlab{}.
\newblock \showarticletitle{A Comparative Analysis of Information Gathering by Chatbots, Questionnaires, and Humans in Clinical Pre-Consultation}. In \bibinfo{booktitle}{\emph{Proceedings of the 2025 CHI Conference on Human Factors in Computing Systems}} \emph{(\bibinfo{series}{CHI '25})}. \bibinfo{publisher}{Association for Computing Machinery}, \bibinfo{address}{New York, NY, USA}, Article \bibinfo{articleno}{639}, \bibinfo{numpages}{17}~pages.
\newblock
\showISBNx{9798400713941}
\urldef\tempurl%
\url{https://doi.org/10.1145/3706598.3713613}
\showDOI{\tempurl}


\bibitem[Lin(2023)]%
        {COPDearlyDetection}
\bibfield{author}{\bibinfo{person}{Ching-Hsiung et~al. Lin}.} \bibinfo{year}{2023}\natexlab{}.
\newblock \showarticletitle{Current Progress of COPD Early Detection: Key Points and Novel Strategies}.
\newblock \bibinfo{journal}{\emph{International journal of chronic obstructive pulmonary disease}}  \bibinfo{volume}{18} (\bibinfo{year}{2023}).
\newblock
\urldef\tempurl%
\url{https://doi.org/doi:10.2147/COPD.S413969}
\showDOI{\tempurl}


\bibitem[Mahmood et~al\mbox{.}(2025)]%
        {Mahmood25}
\bibfield{author}{\bibinfo{person}{Amama Mahmood}, \bibinfo{person}{Shiye Cao}, \bibinfo{person}{Maia Stiber}, \bibinfo{person}{Victor~Nikhil Antony}, {and} \bibinfo{person}{Chien-Ming Huang}.} \bibinfo{year}{2025}\natexlab{}.
\newblock \showarticletitle{Voice Assistants for Health Self-Management: Designing for and with Older Adults}. In \bibinfo{booktitle}{\emph{Proceedings of the 2025 CHI Conference on Human Factors in Computing Systems}} \emph{(\bibinfo{series}{CHI '25})}. \bibinfo{publisher}{Association for Computing Machinery}, \bibinfo{address}{New York, NY, USA}, Article \bibinfo{articleno}{511}, \bibinfo{numpages}{22}~pages.
\newblock
\showISBNx{9798400713941}
\urldef\tempurl%
\url{https://doi.org/10.1145/3706598.3713839}
\showDOI{\tempurl}


\bibitem[Mathews(2023)]%
        {FunctPhysLimitCOPD}
\bibfield{author}{\bibinfo{person}{Anne~M Mathews}.} \bibinfo{year}{2023}\natexlab{}.
\newblock \showarticletitle{The Functional and Psychosocial Consequences of COPD}.
\newblock \bibinfo{journal}{\emph{Respiratory Care}} \bibinfo{volume}{68}, \bibinfo{number}{7} (\bibinfo{year}{2023}), \bibinfo{pages}{914--926}.
\newblock
\urldef\tempurl%
\url{https://doi.org/10.4187/respcare.10542}
\showDOI{\tempurl}


\bibitem[Mayr et~al\mbox{.}(2025)]%
        {Wolfgang}
\bibfield{author}{\bibinfo{person}{Wolfgang Mayr}, \bibinfo{person}{Andreas Triantafyllopoulos}, \bibinfo{person}{Anton Batliner}, \bibinfo{person}{Bj{\"o}rn~W Schuller}, {and} \bibinfo{person}{Thomas~M Berghaus}.} \bibinfo{year}{2025}\natexlab{}.
\newblock \showarticletitle{Assessing the clinical and functional status of COPD patients using speech analysis during and after exacerbation}.
\newblock \bibinfo{journal}{\emph{International Journal of Chronic Obstructive Pulmonary Disease}} (\bibinfo{year}{2025}), \bibinfo{pages}{137--147}.
\newblock


\bibitem[Merkus et~al\mbox{.}(2020)]%
        {Merker20-DEA}
\bibfield{author}{\bibinfo{person}{Julia Merkus}, \bibinfo{person}{Ferdy Hubers}, \bibinfo{person}{Catia Cucchiarini}, {and} \bibinfo{person}{Helmer Strik}.} \bibinfo{year}{2020}\natexlab{}.
\newblock \showarticletitle{Digital Eavesdropper -- Acoustic Speech Characteristics as Markers of Exacerbations in COPD Patients}. In \bibinfo{booktitle}{\emph{Proc.\ RaPID workshop of the 12th LREC}}. \bibinfo{address}{Marseille, France}, \bibinfo{pages}{78}.
\newblock


\bibitem[Milling et~al\mbox{.}(2022)]%
        {SpeechAsBlood2022}
\bibfield{author}{\bibinfo{person}{Manuel Milling}, \bibinfo{person}{Florian~B. Pokorny}, \bibinfo{person}{Katrin~D. Bartl-Pokorny}, {and} \bibinfo{person}{Björn~W. Schuller}.} \bibinfo{year}{2022}\natexlab{}.
\newblock \showarticletitle{Is Speech the New Blood? Recent Progress in AI-Based Disease Detection From Audio in a Nutshell}.
\newblock \bibinfo{journal}{\emph{Frontiers in Digital Health}}  \bibinfo{volume}{4} (\bibinfo{year}{2022}), \bibinfo{pages}{886615}.
\newblock
\urldef\tempurl%
\url{https://doi.org/10.3389/fdgth.2022.886615}
\showDOI{\tempurl}


\bibitem[Milne et~al\mbox{.}(2024)]%
        {Milne2024Sexdifferences}
\bibfield{author}{\bibinfo{person}{Kathryn~M Milne}, \bibinfo{person}{Reid~A Mitchell}, \bibinfo{person}{Olivia~N Ferguson}, \bibinfo{person}{Alanna~S Hind}, {and} \bibinfo{person}{Jordan~A Guenette}.} \bibinfo{year}{2024}\natexlab{}.
\newblock \showarticletitle{Sex-differences in COPD: from biological mechanisms to therapeutic considerations}.
\newblock \bibinfo{journal}{\emph{Frontiers in Medicine}}  \bibinfo{volume}{11} (\bibinfo{date}{mar} \bibinfo{year}{2024}), \bibinfo{pages}{1289259}.
\newblock
\urldef\tempurl%
\url{https://doi.org/10.3389/fmed.2024.1289259}
\showDOI{\tempurl}


\bibitem[Mohamed and maghraby(2014)]%
        {Mohamed14-VCI}
\bibfield{author}{\bibinfo{person}{Enas~Elsayed Mohamed} {and} \bibinfo{person}{Riham Ali~El maghraby}.} \bibinfo{year}{2014}\natexlab{}.
\newblock \showarticletitle{Voice changes in patients with chronic obstructive pulmonary disease}.
\newblock \bibinfo{journal}{\emph{Egyptian Journal of Chest Diseases and Tuberculosis}} \bibinfo{volume}{63}, \bibinfo{number}{3} (\bibinfo{year}{2014}), \bibinfo{pages}{561--567}.
\newblock


\bibitem[Nallanthighal et~al\mbox{.}(2022)]%
        {nallanthighal2022detection}
\bibfield{author}{\bibinfo{person}{Venkata~Srikanth Nallanthighal}, \bibinfo{person}{Aki H{\"a}rm{\"a}}, {and} \bibinfo{person}{Helmer Strik}.} \bibinfo{year}{2022}\natexlab{}.
\newblock \showarticletitle{Detection of COPD exacerbation from speech: comparison of acoustic features and deep learning based speech breathing models}. In \bibinfo{booktitle}{\emph{ICASSP 2022-2022 IEEE International Conference on Acoustics, Speech and Signal Processing (ICASSP)}}. IEEE, \bibinfo{pages}{9097--9101}.
\newblock


\bibitem[Ravanelli et~al\mbox{.}(2021)]%
        {speechbrain}
\bibfield{author}{\bibinfo{person}{Mirco Ravanelli}, \bibinfo{person}{Titouan Parcollet}, \bibinfo{person}{Peter Plantinga}, \bibinfo{person}{Aku Rouhe}, \bibinfo{person}{Samuele Cornell}, \bibinfo{person}{Loren Lugosch}, \bibinfo{person}{Cem Subakan}, \bibinfo{person}{Nauman Dawalatabad}, \bibinfo{person}{Abdelwahab Heba}, \bibinfo{person}{Jianyuan Zhong}, \bibinfo{person}{Ju-Chieh Chou}, \bibinfo{person}{Sung-Lin Yeh}, \bibinfo{person}{Szu-Wei Fu}, \bibinfo{person}{Chien-Feng Liao}, \bibinfo{person}{Elena Rastorgueva}, \bibinfo{person}{François Grondin}, \bibinfo{person}{William Aris}, \bibinfo{person}{Hwidong Na}, \bibinfo{person}{Yan Gao}, \bibinfo{person}{Renato~De Mori}, {and} \bibinfo{person}{Yoshua Bengio}.} \bibinfo{year}{2021}\natexlab{}.
\newblock \showarticletitle{{SpeechBrain}: A General-Purpose Speech Toolkit}.
\newblock \bibinfo{journal}{\emph{arXiv:2106.04624}} (\bibinfo{year}{2021}).
\newblock


\bibitem[S{\o}gaard et~al\mbox{.}(2016)]%
        {sogaard2016incidence}
\bibfield{author}{\bibinfo{person}{Mette S{\o}gaard}, \bibinfo{person}{Morten Madsen}, \bibinfo{person}{Anders L{\o}kke}, \bibinfo{person}{Ole Hilberg}, \bibinfo{person}{Henrik~Toft S{\o}rensen}, {and} \bibinfo{person}{Reimar~W Thomsen}.} \bibinfo{year}{2016}\natexlab{}.
\newblock \showarticletitle{Incidence and outcomes of patients hospitalized with COPD exacerbation with and without pneumonia}.
\newblock \bibinfo{journal}{\emph{International journal of chronic obstructive pulmonary disease}} (\bibinfo{year}{2016}), \bibinfo{pages}{455--465}.
\newblock


\bibitem[Song et~al\mbox{.}(2025)]%
        {ExploreSelf}
\bibfield{author}{\bibinfo{person}{Inhwa Song}, \bibinfo{person}{SoHyun Park}, \bibinfo{person}{Sachin~R Pendse}, \bibinfo{person}{Jessica~Lee Schleider}, \bibinfo{person}{Munmun De~Choudhury}, {and} \bibinfo{person}{Young-Ho Kim}.} \bibinfo{year}{2025}\natexlab{}.
\newblock \showarticletitle{ExploreSelf: Fostering User-driven Exploration and Reflection on Personal Challenges with Adaptive Guidance by Large Language Models}. In \bibinfo{booktitle}{\emph{Proceedings of the 2025 CHI Conference on Human Factors in Computing Systems}} \emph{(\bibinfo{series}{CHI '25})}. \bibinfo{publisher}{Association for Computing Machinery}, \bibinfo{address}{New York, NY, USA}, Article \bibinfo{articleno}{306}, \bibinfo{numpages}{22}~pages.
\newblock
\showISBNx{9798400713941}
\urldef\tempurl%
\url{https://doi.org/10.1145/3706598.3713883}
\showDOI{\tempurl}


\bibitem[Spradley(1979)]%
        {Spradley_P_James_1979}
\bibfield{author}{\bibinfo{person}{James Spradley}.} \bibinfo{year}{1979}\natexlab{}.
\newblock \bibinfo{booktitle}{\emph{The Ethnographic Interview}}.
\newblock \bibinfo{publisher}{Wadsworth,Cengage Learning}.
\newblock
\showISBNx{1-4786-3207-0}


\bibitem[Theopilus et~al\mbox{.}(2025)]%
        {Theopilus25}
\bibfield{author}{\bibinfo{person}{Yansen Theopilus}, \bibinfo{person}{Hilary Davis}, {and} \bibinfo{person}{Sonja Pedell}.} \bibinfo{year}{2025}\natexlab{}.
\newblock \showarticletitle{Digital technology for supporting Aged Care Services: A Scoping Review}. In \bibinfo{booktitle}{\emph{Proceedings of the 2025 CHI Conference on Human Factors in Computing Systems}} \emph{(\bibinfo{series}{CHI '25})}. \bibinfo{publisher}{Association for Computing Machinery}, \bibinfo{address}{New York, NY, USA}, Article \bibinfo{articleno}{506}, \bibinfo{numpages}{12}~pages.
\newblock
\showISBNx{9798400713941}
\urldef\tempurl%
\url{https://doi.org/10.1145/3706598.3713498}
\showDOI{\tempurl}


\bibitem[Triantafyllopoulos et~al\mbox{.}(2024)]%
        {Triantafyllopoulos24-SVF}
\bibfield{author}{\bibinfo{person}{Andreas Triantafyllopoulos}, \bibinfo{person}{Anton Batliner}, \bibinfo{person}{Wolfgang Mayr}, \bibinfo{person}{Markus Fendler}, \bibinfo{person}{Florian Pokorny}, \bibinfo{person}{Maurice Gerczuk}, \bibinfo{person}{Shahin Amiriparian}, \bibinfo{person}{Thomas Berghaus}, {and} \bibinfo{person}{Bj{\"o}rn Schuller}.} \bibinfo{year}{2024}\natexlab{}.
\newblock \showarticletitle{Sustained Vowels for Pre-vs Post-Treatment COPD Classification}. In \bibinfo{booktitle}{\emph{Proceedings of INTERSPEECH}}. \bibinfo{address}{Kos Island, Greece}, \bibinfo{pages}{1410--1414}.
\newblock
\urldef\tempurl%
\url{https://doi.org/10.21437/Interspeech.2024-96}
\showDOI{\tempurl}


\bibitem[Triantafyllopoulos et~al\mbox{.}(2022)]%
        {Triantafyllopoulos22-DBP}
\bibfield{author}{\bibinfo{person}{Andreas Triantafyllopoulos}, \bibinfo{person}{Markus Fendler}, \bibinfo{person}{Anton Batliner}, \bibinfo{person}{Maurice Gerczuk}, \bibinfo{person}{Shahin Amiriparian}, \bibinfo{person}{Thomas Berghaus}, {and} \bibinfo{person}{Björn~W. Schuller}.} \bibinfo{year}{2022}\natexlab{}.
\newblock \showarticletitle{{Distinguishing between pre- and post-treatment in the speech of patients with chronic obstructive pulmonary disease}}. In \bibinfo{booktitle}{\emph{Proceedings of INTERSPEECH}}. \bibinfo{address}{Incheon, South Korea}, \bibinfo{pages}{3623--3627}.
\newblock
\urldef\tempurl%
\url{https://doi.org/10.21437/Interspeech.2022-10333}
\showDOI{\tempurl}


\bibitem[Triantafyllopoulos et~al\mbox{.}(2023)]%
        {Triantafyllopoulos23-H4H}
\bibfield{author}{\bibinfo{person}{Andreas Triantafyllopoulos}, \bibinfo{person}{Alexander Kathan}, \bibinfo{person}{Alice Baird}, \bibinfo{person}{Lukas Christ}, \bibinfo{person}{Alexander Gebhard}, \bibinfo{person}{Maurice Gerczuk}, \bibinfo{person}{Vincent Karas}, \bibinfo{person}{Tobias H{\"u}bner}, \bibinfo{person}{Xin Jing}, \bibinfo{person}{Shuo Liu}, {et~al\mbox{.}}} \bibinfo{year}{2023}\natexlab{}.
\newblock \showarticletitle{HEAR4Health: a blueprint for making computer audition a staple of modern healthcare}.
\newblock \bibinfo{journal}{\emph{Frontiers in Digital Health}}  \bibinfo{volume}{5} (\bibinfo{year}{2023}), \bibinfo{pages}{1196079}.
\newblock
\urldef\tempurl%
\url{https://doi.org/10.3389/fdgth.2023.1196079}
\showDOI{\tempurl}


\bibitem[van Bemmel et~al\mbox{.}(2021)]%
        {van2021automatic}
\bibfield{author}{\bibinfo{person}{Loes van Bemmel}, \bibinfo{person}{Wieke Harmsen}, \bibinfo{person}{Catia Cucchiarini}, {and} \bibinfo{person}{Helmer Strik}.} \bibinfo{year}{2021}\natexlab{}.
\newblock \showarticletitle{Automatic selection of the most characterizing features for detecting COPD in speech}. In \bibinfo{booktitle}{\emph{International Conference on Speech and Computer}}. Springer, \bibinfo{pages}{737--748}.
\newblock


\bibitem[WHO(2024)]%
        {WHO_COPD}
\bibfield{author}{\bibinfo{person}{WHO}.} \bibinfo{year}{2024}\natexlab{}.
\newblock \bibinfo{title}{Chronic obstructive pulmonary disease (COPD)}.
\newblock
\newblock
\newblock
\shownote{Retrieved d. 26-02-2025 Link: \url{https://www.who.int/news-room/fact-sheets/detail/chronic-obstructive-pulmonary-disease-(copd)}}.


\bibitem[Xue et~al\mbox{.}(2025)]%
        {ECG}
\bibfield{author}{\bibinfo{person}{Qiuyue Xue}, \bibinfo{person}{Eric~Steven Martin}, \bibinfo{person}{Jiaqing Liu}, \bibinfo{person}{Ruiqing Wang}, \bibinfo{person}{Antonio Glenn}, \bibinfo{person}{Richard Li}, \bibinfo{person}{Vikram Iyer}, {and} \bibinfo{person}{Shwetak Patel}.} \bibinfo{year}{2025}\natexlab{}.
\newblock \showarticletitle{ECG Necklace: Low-power Wireless Necklace for Continuous ECG monitoring}. In \bibinfo{booktitle}{\emph{Proceedings of the 2025 CHI Conference on Human Factors in Computing Systems}}. \bibinfo{pages}{1--14}.
\newblock


\bibitem[Yuan et~al\mbox{.}(2025)]%
        {yuan2025machine}
\bibfield{author}{\bibinfo{person}{Xiaohan Yuan}, \bibinfo{person}{Chaoqiong Ma}, \bibinfo{person}{Mingzhe Hu}, \bibinfo{person}{Richard~LJ Qiu}, \bibinfo{person}{Elahheh Salari}, \bibinfo{person}{Reema Martini}, {and} \bibinfo{person}{Xiaofeng Yang}.} \bibinfo{year}{2025}\natexlab{}.
\newblock \showarticletitle{Machine learning in image-based outcome prediction after radiotherapy: A review}.
\newblock \bibinfo{journal}{\emph{Journal of Applied Clinical Medical Physics}} \bibinfo{volume}{26}, \bibinfo{number}{1} (\bibinfo{year}{2025}), \bibinfo{pages}{e14559}.
\newblock
\urldef\tempurl%
\url{https://doi.org/10.1002/acm2.14559}
\showDOI{\tempurl}


\end{thebibliography}

\end{document}